# The challenge of listening: The effect of researcher agenda on data collection and interpretation

Rachel E. Scherr, University of Maryland, College Park MD 20742-4111, rescherr@physics.umd.edu
Michael C. Wittmann, University of Maine, Orono ME 04469-5709, wittmann@umit.maine.edu

Even in the relatively favorable circumstances of an individual interview, accurate listening requires careful effort. Our research interests dictate which student statements we attend to during the interview, and which we consider to constitute data during later analysis. Explicit consideration of possible research agendas can increase opportunity for productive research.

## Introduction

As researchers and as instructors, we face the daily challenge of diagnosing student ideas by interpreting their written or spoken statements. Individual student interviews are often considered the gold standard for listening accurately to student ideas. However, even in an open-ended, time-unlimited, one-on-one conversation, accurate listening requires careful effort. We can and do ignore student statements when our own research agenda limits our attention. Explicit consideration of possible research agendas can increase our awareness of the richness of interview data.

The authors of this paper have multiple roles in what follows. One author (MCW) conducted the interview cited here as part of a larger investigation and has since, with collaborators, published a paper describing his findings.[1] The primary author (RES) analyzed the interview transcript some time later. In this paper, MCW is sometimes referred to as "the interviewer," and RES is sometimes referred to as "the observer."

## An Interview on Current and Conductivity

In a previous paper,[1] MCW and his collaborators described a set of interviews in which students were given a simple apparatus (a battery with two leads) to which they could attach steel wire, copper wire, rubber bands, or wood. Students were asked to describe what happens in (*e.g.*) the steel wire when it is connected to the battery terminals.

Both the physics content and the findings of the interviews regarding student models of electrical conduction are described in detail in the cited paper. Our interest here is not in student conceptual understanding of current, but in the negotiation between the interviewer and the subject as they engage in the interview task. For this purpose we cite an excerpt from the beginning of a representative interview and pose the following questions: How is this interview going so far? On what basis do you judge it to be going well or poorly? Does your answer depend on whether you are looking at physics content knowledge, interactions between interviewer and student, or the student's views about knowledge in this topic area?

> Interviewer: So here is the set-up .. battery, your average battery pack, we've got two leads, and let's say that I took a piece of stainless steel, and I placed the stainless steel into the circuit; I made a circuit out of it, right? So attach one end, attach the other end … what happens when you attach both ends?
>
> Student: The circuit loop is complete, so current or electrons will flow out and around back in.
>
> I: Okay. What's going on specifically inside of this piece right here? [indicates steel rod]
>
> S: Electrons are also flowing through that, depending on the makeup of this, I don't know, assuming just…steel, you said… conducting material. I don't know if that's what you…complete conductor, semi-conductor. Do you want me to explain in more detail…?

I: Yeah, sure…go into as much detail as you can. Feel free to make any type of drawings along the way that might help you in your explanation.

S: Okay. I don't know exactly what happens in a resistor, but current is slowed down. It slows down…well, I would assume that there's some kind of chemical property, well I don't know if it's a chemical property, but the oscillating electrons, maybe they're not oscillating as much ...

I: What do you mean by oscillating electrons?

S: Well, that goes back to the first pretest and how the electrons move around the wire, I think the electrons are oscillating, slowly moving around.

I: Okay, so…moving around the circuit?

S: Yeah, that's right. Whatever.

I: […] How are they oscillating? Around what? Can you give me more detail?

S: Mmm…not much! I wouldn't know if they were…well, oscillating perpendicular to the flow or parallel…I don't know, are they oscillating at all?

[Interviewer asks what would be different if the rod were copper.]

S: Well, all these things have uh, these are different chemicals, and they're each going to have their different electrons and protons, so when the electron moves through, maybe there's the attraction and repulsion between this structure, where they're all positive and negative electrons … that's stuff I've never thought about before, and I'm just making it up as we go along!

I: Well, okay! See, that's not bad…after all there's … it's how you come to making it up that's interesting also.[2]

## The effect of observer agenda on data interpretation

We pose again the questions of interest to us in this analysis: How is this interview going so far? On what basis do you judge it to be going well (or poorly)?

The observer (RES) read the entire interview some time after the interview had occurred and initially found the excerpted portion easy to judge: it was not very interesting. What was lacking, in her initial opinion, was information about how the student thinks conduction works. "Sarah" (an alias) offers little sense of a physical mechanism for current; the electrons just 'move around the loop.' She refers to 'oscillation,' but does not say why anything is oscillating, or what the oscillation has to do with current. Worse, the interviewer seems to be having some difficulty getting her to discuss such details; he asks "what she means by oscillating electrons," for example, but her response ("the electrons are oscillating, slowly moving around,") is not clarifying.

Other researchers may agree or disagree with the observer's initial judgment; it does not necessarily tell us the real character of the interview excerpt. What the judgment does tell us is the *nature of the observer's interests.* Her automatic interest is, apparently, in student conceptual understanding of physical mechanism. The observer's interest is only one of a number of primary interests that observers might have (or cultivate). Below we offer an incomplete list of possible research agendas.

1. *Conceptual knowledge of physical mechanism*
   Does the student know the correct physics? Does she have a particular alternative model?
2. *Source of knowledge*
   Is the student's knowledge memorized, constructed, experienced with the senses?
3. *Knowledge construction*
   Is the student skilled at it? How does she generate and select among ideas?
4. *Beliefs about knowledge*
   What does her speech reveal about her epistemological stance?

The observer apparently had item 1 ("Conceptual knowledge of physical mechanism") as her primary research agenda, and since there is relatively little information in the excerpt that is relevant to that interest, it was reasonable for her to judge the interview excerpt as lacking.

An observer with item 2 ("Source of knowledge") as her research agenda, however, might judge the excerpt to be a rich source of information. For example, about halfway through the excerpted portion the student says, "Well, that goes back to the first pretest;" she's drawing on issues raised earlier in her physics class. Later she claims, "we're trying to go to organic chemistry, you know, to draw all these funny pictures," and sketches something resembling a lattice structure. An observer interested in item 3, "Knowledge construction," might be especially interested in the following sequence somewhat later in the interview, in which Sarah explores an apparently new idea:

I: Okay. If I were to put in a different material…rather than steel, I put in copper.
S: Different chemical makeup, so it could …the electrons could either go faster or slower.
I: Okay, what part of the chemical makeup might actually determine that?
[…]
S: Well, all these things have uh, these are different chemicals, and they're each going to have their different electrons and protons, so when the electron moves through, maybe there's the attraction and repulsion between this structure, where they're all positive and negative electrons.

An observer whose interests were primarily epistemological (item 4) would pay special attention when Sarah says, "That's stuff I've never thought about before, and I'm just making it up as I go along!"

We have coded each of Sarah's turns at talk to indicate what sort of information they contain: information about her conceptual knowledge of the physical mechanism, information about her sources of knowledge, and so on.[3] We found that within Sarah's first eighteen turns at talk, Sarah made nineteen utterances that were codable according to the four categories described. Of those nineteen utterances, six were relevant to describing a physical mechanism, five to sources of knowledge, five to knowledge construction, and four to epistemological issues. To pay attention to only one of these four categories is to fail to hear much of what the student is saying.

Our coding scheme is nontrivial to execute, and we are still examining issues of inter-rater reliability (our coding represents consensus after discussion). Our intention, however, is not to claim that we have definitively identified the nature of each of Sarah's statements. Instead, we intend only to illustrate that different research agendas can result in very different judgments of the interview excerpt. In particular, *an observer's interests dictate which student statements are considered to constitute data.* The observer's initial judgment – that the interview excerpt did not contain much information – reveals that she did not initially attend to agendas other than #1 (conceptual knowledge of physical mechanism). Her interests acted as a "filter" on the data.

## The effect of interviewer agenda on data collection

The observation that an observer's research agenda may affect data interpretation naturally raises questions about the research agenda of the *interviewer*. It is reasonable to assume that the interviewer has a "filter" as well, and that his interpretation of student statements during the interview shapes the course of the conversation to some extent. Because research agendas are often implicit, it is rarely sufficient to simply ask the interviewer about his intentions. Instead, we examine the transcript for clues to his interests.

We have coded the first eighteen interviewer turns at talk to indicate what sort of information the interviewer was requesting of the student. The great majority of the codable prompts (12/14) are for conceptual knowledge of physical mechanism. For example, the interviewer asks, "What do you mean by oscillating electrons? …How are they oscillating? Around what?" There are two prompts for knowledge construction (*e.g.,* "It's how you come to making it up that's interesting also"), but the interviewer's primary interest appears to be in the first of the four areas cited above.

Additional evidence for this identification appears in the published paper describing these interviews. In that article, the authors represent Sarah in the following way:

> "Sarah first described atomic lattice vibrations in a heated wire impeding electron flow, but then changed her response to say that the energy, when transferred to electrons, helped the electron flow."

The authors make other comments about Sarah that indicate they were also paying attention to research agenda #3 (knowledge construction). However, their primary interest appears to lie in Sarah's conceptual model for electrical conduction.

The effect of the interviewer's research agenda on data collection appears to be similar to the effect of the observer's agenda on data interpretation: the interviewer shows selective attention to particular student statements. Consider, for example, the following exchange:

S: Okay. I don't know exactly what happens in a resistor, but current is slowed down. It slows down…well, I would assume that there's some kind of chemical property, well I don't know if it's a chemical property, but the oscillating electrons, maybe they're not oscillating as much ...
I: What do you mean by oscillating electrons?

Sarah's statement includes both hints of a conceptual model ("oscillating electrons") and information regarding the sources of her knowledge ("I don't know exactly what happens…there's some kind of a chemical property"). The interviewer might make any one of several possible moves at this point; for example, he might ask Sarah what she means by a chemical property, or whether it's legitimate to speculate about mechanisms one doesn't remember exactly. Instead, though, he asks for more detail about the oscillations. His choice of response calls our attention to his primary interest in the conceptual content of Sarah's ideas. And, presumably, his response calls the *student's* attention to his interests also, as she and he together negotiate the course of the interview. His choice of response encourages her to provide more conceptual information in the future.

## Discussion

We hope to have shown in this paper that hearing all of what students are saying requires careful effort. We are not automatically conscious of everything a student says; the "filter" of our own research interests blocks some student statements. Such a filter is not inappropriate for researchers; specialization is usually necessary for detailed analysis. However, to the extent that our research agendas are unexamined, they may control our attention inappropriately. Conscious consideration of possible research agendas widens the range of opportunity for productive research.